# Clock Distributing for BaF$_2$ Readout Electronics at CSNS-WNS [*]

Bing He (何兵)[1,2]    Ping Cao (曹平)[1,2*]    De-Liang-Zhang (张德良)[1,3]
Qi Wang (王奇)[1,3]    Ya-Xi Zhang (张雅希)[1,3]    Xin-Cheng Qi (齐心成)[1,2]    Qi An (安琪)[1,2]

[1] State Key Laboratory of Particle Detection and Electronics, University of Science and Technology of China, Hefei 230026, China
[2] School of Nuclear Science and Technology, University of Science and Technology of China, Hefei 230026, China
[3] Department of Modern Physics, University of Science and Technology of China, Hefei 230026, China

Abstract: BaF$_2$ (Barium Fluoride) detector array is designed for the measurement of (n,γ) cross section precisely at CSNS-WNS (white neutron source at China Spallation Neutron Source). It is a 4 π solid angle-shaped detector array consisting of 92 BaF$_2$ crystal elements. To discriminate signals from BaF$_2$ detector, pulse shape discrimination methodology is used, which is supported by waveform digitization technique. There are total 92 channels for digitizing. The precision and synchronization of clock distribution restricts the performance of waveform digitizing. In this paper, the clock prototype for BaF$_2$ readout electronics at CSNS-WNS is introduced. It is based on PXIe platform and has a twin-stage tree topology. In the first stage, clock is distributed from the tree root to each PXIe crate synchronously through coaxial cable over long distance, while in the second stage, clock is further distributed to each electronic module through PXIe dedicated differential star bus. With the help of this topology, each tree node can fan out up to 20 clocks with 3U size. Test result shows the clock jitter is less than 20ps, which can meet the requirement of BaF$_2$ readout electronics. Besides, this clock system has advantages of high density, simplicity, scalability and cost saving, which makes it can be used in other applications of clock distributing preciously.

## 1 Introduction

CSNS-WNS is a scientific facility in Dongguan, it can provide intense flux, good energy spectrum and great resolution neutrons [1]. BaF$_2$ detector array (Fig.1) is a 4 π angle-shaped detector array consisting of 92 crystal elements in WNS. Its main target is to measure the cross section of (n,γ) reaction. To discriminate the signals form BaF$_2$ detector, pulse shape discrimination methodology is used, which is supported by waveform digitization technique based on high performance ADC [2]. To deal with the massive data produced, the readout electronics is distributed in 4 PXIe crates, and a clock system is also demanded to provide low jitter and low skew clocks. PXIe platform can bring great advantages for the design of readout electronics. However, it also brings challenge for the clock distributing. Traditionally, clocks are fanned out to modules through fiber or differential cable from the front panel of readout module. However, these schemes would meet difficulty introduced by the limited 3U size of PXIe module.  In this paper, a distributed clock system based on PXIe crates is proposed. It can transmit high-quality clock through long coaxial cable between crates. Besides, precise clock can also be distributed to each slot in crate without any extra cables. Furthermore, this clock system has advantages of scalable, universally applicable, space-saving and cost-saving ability [3] [4].

Generally, clock is distributed from a global node and fanned out to each receiving slave node. For the end-cap time of flight (TOF) upgrade of Beijing Spectrometer (BESIII), two types of VME 9U module are used to distribute clock. One is assigned as the master clock module and another is slave clock module. Master and slave modules are located in two VME crates respectively. Master clock module transmits clock to slave clock module by fiber because of the long distance between crates. Furthermore, once received by slave module, clock is distributed to all readout electronic modules in the same crate through differential cables from the front panel of slave module [5] [6] [7].

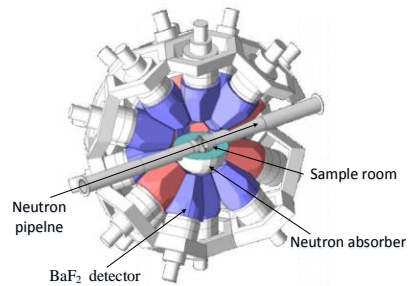

Fig. 1. BaF2 detector array

In the case of nTOF experiment in CERN, the sampling clocks are generated by a central clock generator and are distributed to digitizer modules via the fan-out units (integral parts of the generator) by equal length cables. These digitizer modules are located in 19 CPCI crates. Care has been taken to assure that the skew of each output is less than 200ps in respect to all other outputs [8].

[*] Supported by NSAF (U1530111) and National Natural Science Foundation of China (11005107).
1) E-mail: cping@ustc.edu.cn

As for the clock system of GTAF (gamma-ray total absorption facility) in Chinese Institute of Atomic Energy, All commercial digitizers (acqiris DC271A) are located into one CPCI crate. The reference clock is transmitted to one digitizer at first. Then this digitizer distributes clocks further to other digitizers via ASBus that can distribute clock signals along a plug-in front panel bus. Up to 7 digitizers can be synchronized with the ASBus [9].

For the case of $BaF_2$ detector array at CSNS-WNS, readout electronics is designed based on PXIe platform because of its advantages of high data transmission capacity. The up to 8GB/s of signal-slot transfer bandwidth can guarantee ability of uploading massive data produced by waveform digitizing to crate controller in real-time and in parallel. However, clock distributing meets design challenges caused by the size of PXIe module. The 3U euro PXIe module is about 90 mm, but it should fan out at least 18 channels for the case of $BaF_2$ readout electronics. Moreover, even more serious is that trigger distributing also needs module panel space. So there is no space for photoelectric converters or differential cable connectors if traditional clock distributing methodology is used. To solve these problems, a high-density clock distributing system based on PXIe platform is proposed in this paper.

## 2 Architecture of the clock system

The clock distribution network has a tree topology as illustrated in Fig.2. Global clock module is appointed as the root of this tree, while each PXIe module is appointed as a clock-receiving node in this tree.

The clock distribution tree has a twin-stage structure. In the first stage, global clock module distributes clock to local clock modules in each PXIe crate. The clock loopback is used to correct the phase error that is caused by different transmission path. In the second stage, clock modules will fan out clocks to other modules in the same crate, acting as a buffer.

In addition, in consideration of the compatibility with PXIe backplane clocking resource [10], the frequency of the clock system is 100MHZ. Actually, the 12bit ADC for waveform digitization runs at sampling rate of 1000MSps, which need reference clock with good specification of low jitter and low skew. In order to ensure the accuracy of clock distributing, the clock jitter and clock skew should be less than 100ps. However, in practice a few design challenges are involved into this approach:

▪ Transmit clocks from global clock module to local clock modules over long distance with limited panel space. The clock skew and jitter should be calibrated at the same time.

▪ Distribute clocks to all 16 slots in a crate through a 3U size of clock module with high quality.

## 3 Clock system implementation

### 3.1 Clock distributing between crates

For the clock distribution between crates, fibers and differential cables are not appropriate because of the shortage of panel space. So coaxial cable should be used to transmit clocks to local clock modules, and micro-miniature coaxial (MMCX) connectors are used to save panel space further. But coaxial cable is bound to introduce another problem, the clock signals will be attenuated and poor clocks will be received, especially for long distance. So new technique should be used to solve this difficulty.

Equalization can compensate the distortion caused by signal attenuation. It is critical to improve signal quality over long distance [11]. So equalizers can be used in this clock system to improve the clock quality received by local clock modules. Driver and equalizer should be considered as a pair of signal transmission. They can extend the valid transmission distance over long single-ended media. The driver is a high-speed differential buffer with adjustable output amplitude, while the equalizer is optimized for equalizing signal transmitted over balanced copper cable. However, the lowest working frequency of the equalizer is 150MHZ, so the clock transmitted between crates should be upgraded to 200MHZ instead. The clock link via coaxial cable is shown in Fig.3 [12].

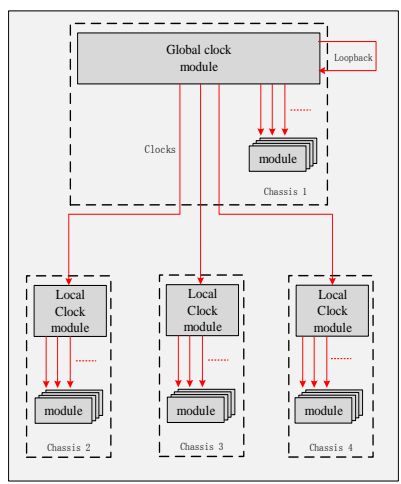

Fig. 2. Clock distribution tree

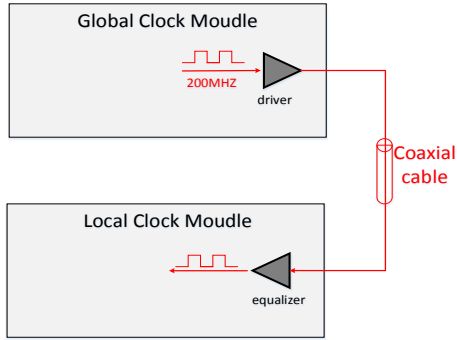

Fig. 3. Clock distribution via coaxial cable

As discussed above, clock skew is also a critical factor for clock distributing. Clock skew introduced by the length difference of cables should be eliminated or calibrated. The simplest way to eliminate skew is to make the length of each distributing cable equals to each other. But this manner is troublesome and time consuming. Actually, PLL with adjustable delay can be used to correct the clock skew. It is convenient and high-efficient to compensate the measured skew by reconfiguring the clock delay. The PLL has ability of analog delay adjustment with step of 25ps, which can achieve high synchronization accuracy.

### 3.2 Clock distributing inside crate

To distribute clocks to 16 slots in a 3U PXIe crates, fanning out by extra cables is not suitable because of the shortage of panel space. A scheme to distribute high-quality clocks without use of cables should be adopted.

PXIe crate provides three kinds of slot, a system controller slot, a system timing slot (STM) and 16 peripheral slots. It also provides rich backplane traces with high performance for trigger and synchronization. The PXIe_DSTARA is a differential star trigger bus designed to distribute point-to-point, high-quality LVPECL clocks from the STM slot to each peripheral slot (Fig.3). The maximum clock skew of the two signals within PXIe_DSTARA is less than 25ps, while the jitter introduced to the clock is less than 5 ps.

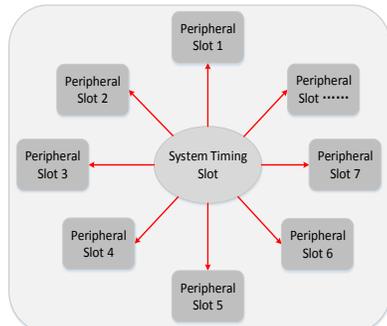

Fig. 3. PXIe_DSTARA topology

So PXIe_DSTARA can be used to distribute clocks to each slot inside crate. Time module is located in STM slot, while digitizing modules are located in peripheral slots. Based on this scheme, up to 16 clocks can be distributed synchronously over the differential star trigger bus. Furthermore, this scheme has advantage of convenient, concise, space saving and meeting the quality requirement of the readout electronics as well.

## 4 Experiment and Verification

### 4.1 Test platform

According to this proposed clock distributing structure, for the purpose of evaluating clock quality, a test platform has been constructed with a global clock module and a PXIe crate shown in Fig.4. The global clock module (shown in Fig. 5) is located into the STM slot of crate though the backplane connector. In the global clock module, PLL would produce high-quality 100M and 200M LVDS clocks sourcing form the oscillator. Then the 200MHZ clocks are fanned out through cables, while the 100MHZ clocks are distributed through PXIe_DSTARA on backplane.

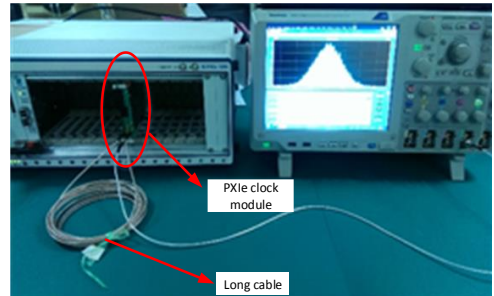

(1)Picture of Platform

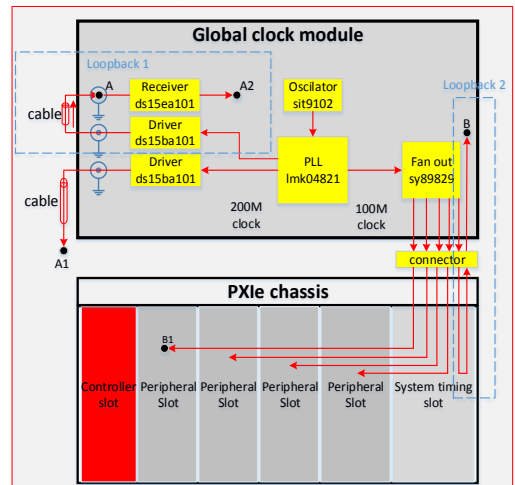

(2)Diagram of Platform
Fig. 4.Test platform

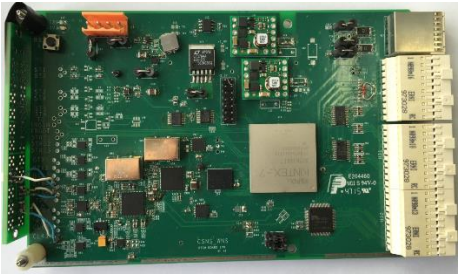

Fig. 5. Global clock module

As illustrated in Fig.4, two loopbacks are used to simulate the clock distribution between and within crates. There are five test points (A, A1, A2, B, B1) used to measure the quality of this clock distributing system.

The loopback 1 simulates the clock distribution process between crates. The clock can be transmitted back to global clock module through long coaxial cable over this loopback channel. In Fig.4, signal A denotes the single-ended clock received by local clock module. Signal A2 represents for the differential clock after being equalized. To evaluate clock skew between crates, signal of A should be compared with signal of A1 denoting clock distributed to other crates.

In the loopback 2, the clocks would be fanned out with LVPECL signaling technique to each slot though the dedicated PXIe_DSTARA backplane bus. One PXIe_DSTARA would trace back to the global clock module through backplane. The signal of B is the clock received by each slot. For the clock skew measurement, the skew between clock B and clock B1 should be measured.

### 4.2 Test result

All data about the quality of the clock distributing system is measured and acquired by the Tektronix DPO5104 oscilloscope. The jitter of the distributed clock is characterized in TIE (Time Interval Error) at a population of $10^5$, which is enough to get a highly accurate statistical result.

Test results of jitter are shown in Fig.6 from which two conclusions can be drawn. The first conclusion is that the clock jitter of A2 and B is less than 20ps, which can meet the requirement of the readout electronics. The second one is that the clock jitter has been corrected from 30ps to 10ps by the equalizer, which demonstrates that distributing clock over coaxial cable combined with equalizer can achieve high-quality clock distributing over long distance.

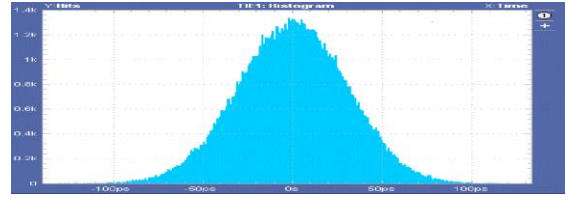

(1)Jitter of A, TIE=30ps

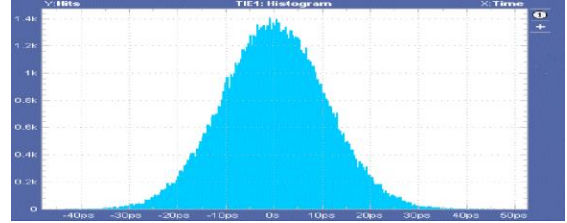

(2)Jitter of A2, TIE=10ps

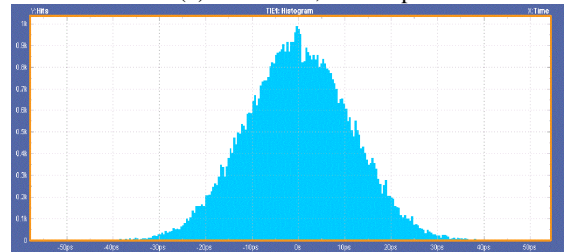

(3)Jitter of B, TIE=12ps

Fig. 6. Clock jitter

## 5 Conclusion

An optimized clock distributing system with high performance and density has been designed for CSNS-WNS BaF$_2$ readout electronics. Long distance of coaxial cable, equalizer and PXIe dedicated differential bus are used to distribute clocks synchronously and precisely. Test result shows that the clock jitter is less than 20ps that can meet the requirements of the readout electronics. It is a breakthrough that only 4 PXIe modules are used to distribute clocks to 68 modules. Compared with traditional clock distributing manner, this clock system is scalable, universally applicable, space and cost saving. This synchronous clock distribution method can also be used in other applications of clock distributing precisely.

**Acknowledge:**
*The authors gratefully acknowledge all members of CSNS-WNS for their earnest support and help during the work.*